\documentclass[]{svjour2}

\smartqed

\usepackage{amssymb,amsmath}
\usepackage{graphicx}

\journalname{J Stat Phys}

\begin{document}

\title{Short-Distance Symmetry of Pair Correlations in
Two-Dimensional Jellium}

\titlerunning{Short-Distance Symmetry of Pair Correlations in 2D Jellium}

\author{Ladislav \v{S}amaj}

\institute{Institute of Physics, Slovak Academy of Sciences, 
D\'ubravsk\'a cesta 9, SK-84511 Bratislava, Slovakia \\
\email{Ladislav.Samaj@savba.sk}}

\date{Received:  / Accepted: }

\maketitle

\begin{abstract} 
We consider the two-dimensional one-component plasma (jellium) of mobile 
pointlike particles with the same charge $e$, interacting pairwisely by 
the logarithmic Coulomb potential and immersed in a fixed neutralizing 
background charge density.
Particles are in thermal equilibrium at the inverse temperature $\beta$,
the only relevant dimensionless parameter is the coupling constant
$\Gamma\equiv\beta e^2$. 
In the bulk fluid regime and for any value of the coupling constant 
$\Gamma=2\times{\rm integer}$, \v{S}amaj and Percus 
[J. Stat. Phys. {\bf 80}, 811--824 (1995)] have derived an infinite sequence
of sum rules for the coefficients of the short-distance expansion of
particle pair correlation function.
In the context of the equivalent fractional quantum Hall effect, by using
specific methods of quantum geometry Haldane
[PRL {\bf 107}, 116801 (2011) and arXiv:1112.0990v2] derived a self-dual
relation for the Landau-level guiding-center structure factor.
In this paper, we establish the relation between the guiding-center structure 
factor and the pair correlation function of jellium particles. 
It is shown that the self-dual formula, which provides an exact relation 
between the pair correlation function and its Fourier component, 
comes directly from the short-distance symmetry of the bulk jellium. 
The short-distant symmetry of pair correlations is extended to the
semi-infinite geometry of a rectilinear plain hard wall with a fixed surface 
charge density, constraining particles to a half-space. 
The symmetry is derived for the original jellium model as well as its 
simplified version with no background charge (charged wall surface with 
``counter-ions only'').
The obtained results are checked at the exactly solvable free-fermion coupling
$\Gamma=2$.  
  

\end{abstract}

\renewcommand{\theequation}{1.\arabic{equation}}
\setcounter{equation}{0}

\section{Introduction} \label{Sect1}
The study of statistical mechanics of classical (i.e. non-quantum) systems of 
particles interacting pairwisely via the Coulomb potential is of primary 
interest in many branches of condensed matter and soft matter physics.

In Gauss units and with the vacuum dielectric constant $\varepsilon=1$, 
the Coulomb potential $\phi$ at point ${\bf r}$ of an infinite Euclidean space 
of dimension $d$, induced by a unit charge at the origin ${\bf 0}$,
is defined as the solution of the $d$-dimensional Poisson equation 
\begin{equation}
\Delta \phi({\bf r}) = - s_d \delta({\bf r}) 
\end{equation}
supplemented with the boundary condition of vanishing electric field at 
infinity.
Here, $s_d = 2\pi^{d/2}/\Gamma(d/2)$ ($\Gamma$ denotes the Gamma function
\cite{Gradshteyn}) is the surface area of the $d$-dimensional unit sphere. 
In particular,
\begin{equation}
\phi({\bf r}) = \left\{
\begin{array}{ll}
-\ln(r/L) \phantom{aaa} &  \mbox{if $d=2$,} \cr & \cr
\displaystyle{\frac{r^{2-d}}{d-2}} & \mbox{otherwise,}
\end{array} \right.
\end{equation}
where $r=\vert {\bf r}\vert$ and a free length scale $L$ fixes the zero
of the two-dimensional (2D) Coulomb potential; for simplicity, we set $L=1$.
The Fourier component of such potential exhibits the characteristic
form $1/k^2$ with singularity at $k=0$ which keeps many generic properties 
of three-dimensional (3D) Coulomb systems with $1/r$ interaction potential.
The interaction energy of two charges $q$ and $q'$ at the respective positions 
${\bf r}$ and ${\bf r}'$ is given by $q q' \phi(\vert {\bf r}-{\bf r}'\vert)$.  
The 2D Coulomb system can be represented as parallel infinite charged lines 
interacting in 3D which are perpendicular to the given surface and as such 
are of practical interest in the field of polyelectrolytes.

From among numerous types of Coulomb models we shall concentrate on
the so-called one-component plasma (OCP), or jellium, which represents 
a reasonable simplification of realistic systems of atomic nuclei and electrons.
Jellium consists of mobile pointlike particles with equivalent (say
elementary) charge $e$, immersed in a uniform neutralizing background
charge density. 
The system is considered to be in thermal equilibrium at the temperature $T$,
or the inverse temperature $\beta=1/(k_{\rm B}T)$.
As for any Coulomb system, its thermal equilibrium is exactly solvable in the 
high-temperature region within the linear Debye-H\"uckel or nonlinear
Poisson-Boltzmann mean-field theories \cite{Baus80}.
The long-range tail of the Coulomb potential implies exact constraints
(sum rules) on the moments of bulk particle correlation functions, 
see review \cite{Martin88}.
In any spatial dimension, these sum rules include the zeroth- and second-moment 
Stillinger-Lovett conditions \cite{Stillinger68a,Stillinger68b} and the
fourth-moment (compressibility) condition 
\cite{Baus78,Vieillefosse75,Vieillefosse85}.
In 2D, the sixth-moment condition was derived in Ref. \cite{Kalinay00}.
Another kind of sum rule, relating two lowest-order coefficients of
the short-distance expansion of the correlation function was derived for 
the 3D jellium by Jancovici \cite{Jancovici77}.

In this paper, we shall concentrate on the 2D OCP.
Its thermodynamics and particle correlation functions depend only on 
the coupling constant $\Gamma=\beta e^2$, the particle density scales 
appropriately distances. 
Besides the high-temperature Debye-H\"uckel limit $\Gamma\to 0$, 
the 2D jellium is exactly solvable also at $\Gamma=2$ by mapping onto 
free fermions \cite{Alastuey81,Jancovici81}.
In the bulk regime, the two-body correlation functions decay exponentially
at asymptotically large distances for $\Gamma\to 0$ while the decay is
Gaussian at $\Gamma=2$.    
The solvable cases involve also semi-infinite \cite{Jancovici82} or fully 
finite geometries, see reviews \cite{DiFrancesco94,Forrester98,Jancovici92}. 

The extension of the 3D relation between the first two coefficients of
the short-distance expansion of the correlation function \cite{Jancovici77}
to 2D is nontrivial.
For a sequence of the coupling constants $\Gamma = 2\times{\rm integer}$, 
using a mapping of the 2D jellium onto a one-dimensional many-component 
anticommuting field theory on a discrete chain, a symmetry of the bulk 
two-body correlations with respect to a complex transformation of particle 
coordinates leads to a functional relation which implies an infinite sequence 
of the relations among the coefficients of the short-distance expansion of 
the correlation function \cite{Samaj95}.
Since only every second relation of the sequence is effective, this result 
does not provide an explicit form of the correlation function but strongly 
restricts its possible functional forms.      
The short-distance symmetry was extended to multi-particle correlations
in Ref. \cite{Samaj98}.

In the fractional quantum Hall effect \cite{Prange87}, the Hall conductance
exhibits plateaux indexed via filling fractions of Landau levels.
The partition function of the 2D OCP is formally the normalization factor  
of Laughlin's proposal of the wave function \cite{DiFrancesco94}.
In the context of the fractional quantum Hall effect, using specific methods 
of quantum geometry Haldane derived a self-dual relation for the Landau-level 
guiding-center structure function \cite{Haldane11a,Haldane11b}. 

In this paper it is shown that the guiding-center structure function is
related to the two-body density of the 2D jellium. 
For any coupling $\Gamma = 2\times{\rm integer}$, the counterpart of 
the self-dual formula comes directly from the short-distance
symmetry of the pair correlations for the bulk 2D jellium.
The self-dual formula relates the pair correlation function and 
its Fourier component. 
Another new result is the extension of the short-distant symmetry of pair 
correlations to the semi-infinite geometry of a charged rectilinear 
plain hard wall (line), constraining charged particles to a half-space. 
The extension is worked out for the original 2D jellium as well as its 
simplified version with no background charge, namely the charged line
with ``counter-ions only''.
The obtained results are checked at the exactly solvable free-fermion 
coupling $\Gamma=2$.

The paper is organized as follows.
In Sect. \ref{Sect2} we present within the canonical ensemble the definition of 
thermodynamic quantities of the 2D OCP constrained to an arbitrary domain 
and derive the general symmetry relation for two-body densities.
The short-distance symmetry of the pair correlation function in the bulk regime
is recapitulated in Sect. \ref{Sect3}.
Sect. \ref{Sect4} deals with the consequences of the studied symmetry
in the Fourier space.
The correlation-function counterpart of the guiding-center structure factor 
from the fractional quantum Hall effect is defined and the self-dual
relation between its Fourier and Euclidean pictures is derived.
In Sect. \ref{Sect5}, the short-distant symmetry of pair correlations 
is extended to the semi-infinite geometry of a rectilinear plain hard wall 
charged by a fixed line charge density. 
The original jellium model as well as its simplified version with 
no bulk background charge density (system with surface charge density
and ``counter-ions only'') are studied.
The obtained results, valid for any $\Gamma = 2\times{\rm integer}$, 
are checked at the exactly solvable free-fermion coupling $\Gamma=2$.  
A short recapitulation is given in the concluding Sect. \ref{Sect6}. 

\renewcommand{\theequation}{2.\arabic{equation}}
\setcounter{equation}{0}

\section{General symmetry} \label{Sect2}
Let $N$ pointlike mobile particles $j=1,2,\ldots,N$ of charge $e$ be 
constrained to a 2D domain $\Lambda$ of surface $\vert\Lambda\vert$ 
by plain hard walls (lines) which are located at the domain boundary 
$\partial\Lambda$ and may carry a uniform line-charge density
$e \sigma$ ($\sigma=0$ in the case of neutral boundaries).
The domain points ${\bf r}=(x,y)$ will be often written in the complex notation
\begin{equation}
z = x + {\rm i} y , \qquad \bar{z} = x - {\rm i} y .
\end{equation}
A background charge density $\rho_b$ is distributed uniformly over $\Lambda$.
The condition of overall charge neutrality reads as
\begin{equation}
\rho_b \vert\Lambda\vert + N e + \sigma \vert\partial\Lambda\vert e = 0 .
\end{equation}
In this paper, we consider only infinite and semi-infinite domains with 
$N\to\infty$ for which it holds that
$\vert\partial\Lambda\vert/\vert\Lambda\vert\to 0$.
Consequently, $\rho_b = -e n$ with $n=N/\vert\Lambda\vert$ being 
the mean density of particles.

The dielectric constant of the walls $\varepsilon_W$ is considered
to be the same as the one $\varepsilon$ of the medium in which the particles 
are immersed, say $\varepsilon_W = \varepsilon = 1$, so there are no
image forces.  
Two particles at positions ${\bf r}$ and ${\bf r}'$ interact by 
the 2D Coulomb energy $-e^2 \ln\vert {\bf r}-{\bf r}'\vert$.
Let the electrostatic potential induced by uniform surface $\rho$
and line $\sigma$ charge densities at position $(z,\bar{z})\in\Lambda$
be denoted by $v(z,\bar{z})$; the corresponding one-body Boltzmann
factor at the inverse temperature $\beta$ is 
$w(z,\bar{z})\equiv \exp[-\beta v(z,\bar{z})]$. 
 
Due to the presence of the rigid neutralizing background, the jellium system 
is studied in the canonical ensemble. 
The partition function is given by
\begin{equation}
Z_N = \frac{1}{N!} \int_{\Lambda} \prod_{j=1}^N {\rm d}^2r_j w(z_j,\bar{z}_j)
\prod_{(j<k)=1}^{N} \vert z_j-z_k\vert^{\Gamma} ,
\end{equation}
where $\Gamma=\beta e^2$ is the coupling constant.
We study the special cases of $\Gamma=2\gamma$ where $\gamma=1,2,\ldots$ 
is a (positive) integer. 
The canonical averaging is defined as 
\begin{equation}
\langle \cdots \rangle = \frac{1}{Z_N} \frac{1}{N!} 
\int_{\Lambda} \prod_{j=1}^N {\rm d}^2r_j w(z_j,\bar{z}_j)
\prod_{(j<k)=1}^{N} \vert z_j-z_k\vert^{\Gamma} \cdots .
\end{equation}
The microscopic total number density of particles at point ${\bf r}$ is 
given by $\hat{n}({\bf r}) = \sum_{j=1}^N \delta({\bf r}-{\bf r}_j)$.
At the one-particle level, one defines the average number density
\begin{equation}
n({\bf r}) = \langle \hat{n}({\bf r}) \rangle . 
\end{equation}
At the two-particle level, one introduces the two-body density
\begin{equation} \label{twobody}
n^{(2)}({\bf r},{\bf r}') = \left\langle \sum_{j\ne k} \delta({\bf r}-{\bf r}_j)
\delta({\bf r}'-{\bf r}_k) \right\rangle =
\langle \hat{n}({\bf r}) \hat{n}({\bf r}') \rangle
- n({\bf r}) \delta({\bf r}-{\bf r}') 
\end{equation}
and the (truncated) pair correlation function
\begin{equation}
h({\bf r},{\bf r}') = \frac{n^{(2)}({\bf r},{\bf r}')}{n({\bf r})n({\bf r}')}
-1  
\end{equation}
which vanishes at asymptotically large distances 
$\vert {\bf r}-{\bf r}'\vert\to\infty$.

To derive the general symmetry of interest for two-body densities, 
in analogy with Ref. \cite{Samaj95} we write down explicitly their integral 
representation from the definition (\ref{twobody}):
\begin{eqnarray}
\frac{n^{(2)}({\bf r}_1,{\bf r}_2)}{w({\bf r}_1) w({\bf r}_2)
\vert {\bf r}_1-{\bf r}_2 \vert^{2\gamma}} & = & \frac{1}{2 Z_N} \frac{1}{(N-2)!}
\int_{\Lambda} \prod_{j=3}^N {\rm d}^2r_j w(z_j,\bar{z}_j) \nonumber \\ 
& & \times \vert z_1-z_j \vert^{2\gamma} \vert z_2-z_j \vert^{2\gamma} 
\prod_{(j<k)=3}^{N} \vert z_j-z_k\vert^{2\gamma} . \label{repre} 
\end{eqnarray}
For every particle index $j=3,\ldots,N$, the product 
$\vert z_1-z_j \vert^2 \vert z_2-z_j \vert^2 = 
(z_1-z_j) (\bar{z}_1-\bar{z}_j) (z_2-z_j) (\bar{z}_2-\bar{z}_j)$
is invariant with respect to the following transformation 
of particle coordinates
\begin{equation} \label{transcoor}
z'_1 = z_2 , \qquad \bar{z}'_1 = \bar{z}_1 , \qquad 
z'_2 = z_1 , \qquad \bar{z}'_2 = \bar{z}_2 .
\end{equation}
In the center-of-mass basis 
\begin{equation} \label{mass}
{\bf R} = \frac{1}{2} \left( {\bf r}_1 + {\bf r}_2 \right) ,
\qquad {\bf r} = {\bf r}_1 - {\bf r}_2 ,
\end{equation}
this transformation takes the form
\begin{equation}
{\bf R}' = {\bf R} , \qquad {\bf r}' = {\rm i} {\bf z}\times {\bf r} ,
\end{equation}
where ${\bf z}={\bf x}\times {\bf y}$ is the unit vector perpendicular
to the considered $(x,y)$ plane. 
Since the rhs of Eq. (\ref{repre}) is invariant with respect to 
the transformation (\ref{transcoor}), one arrives at the symmetry relation
\begin{eqnarray} 
& \displaystyle{\frac{n^{(2)}(z_1,\bar{z}_1,z_2,\bar{z}_2)}{w(z_1,\bar{z}_1) 
w(z_2,\bar{z}_2) \left[ (z_1-z_2) (\bar{z}_1-\bar{z}_2)\right]^{\gamma}}} 
\nonumber \\ & \qquad \qquad = 
\displaystyle{\frac{n^{(2)}(z_2,\bar{z}_1,z_1,\bar{z}_2)}{w(z_2,\bar{z}_1) 
w(z_1,\bar{z}_2) \left[ (z_2-z_1) (\bar{z}_1-\bar{z}_2)\right]^{\gamma}}} , 
\label{final1}  
\end{eqnarray}
where the two-body function $w(z_j,\bar{z}_k)$ with $j\ne k$ is the obvious 
generalization of the one-body Boltzmann factor 
$w(z\to z_j,\bar{z}\to \bar{z}_k)$.
The relation (\ref{final1}) can be simplified to the one
\begin{equation} \label{final2}
\frac{n^{(2)}(z_1,\bar{z}_1,z_2,\bar{z}_2)}{w(z_1,\bar{z}_1) w(z_2,\bar{z}_2)} 
= (-1)^{\gamma}
\frac{n^{(2)}(z_2,\bar{z}_1,z_1,\bar{z}_2)}{w(z_2,\bar{z}_1) w(z_1,\bar{z}_2)} .
\end{equation}
The practical realization of this symmetry formula depends on the form of
$\Lambda$-domain, which manifests itself via the specific form of 
the one-body Boltzmann factor $w(z,\bar{z})$ and the coordinate dependence 
of the two-body density.

\renewcommand{\theequation}{3.\arabic{equation}}
\setcounter{equation}{0}

\section{Bulk regime} \label{Sect3}
For an infinite Euclidean surface, the mean density of particles is
constant, $n({\bf r}) = n$ and $\rho_b=-e n$, so the electroneutrality
constraint is local. 
Let us assume a circular dependence of the electrostatic potential induced by 
the homogeneous (infinite) background, i.e. the corresponding one-body energy 
of particles is given by the radial Poisson equation
\begin{equation}
\frac{1}{r} \frac{{\rm d}}{{\rm d}r} \left[ r \frac{{\rm d}u(r)}{{\rm d}r} 
\right] = 2\pi n e .
\end{equation}
The solution of this differential equation is $u(r) = \pi n er^2/2$,
so the one-body Boltzmann factor reads as
\begin{equation} \label{Bf}
w(z,\bar{z}) = \exp\left( -\frac{1}{2} \Gamma \pi n r^2 \right) 
= \exp\left( -\gamma \pi n z\bar{z} \right) .
\end{equation}
The particle density $n$ is the parameter which only scales appropriately
the distances.
We define the length
\begin{equation} \label{lengtha}
a = \frac{1}{\sqrt{\gamma\pi n}}
\end{equation}
and express all distances in units of $a$; in those units, $n=1/(\pi\gamma)$.
The two-body generalization of the one-body Boltzmann factor 
$w(z,\bar{z}) = \exp(-z\bar{z})$ reads as
\begin{equation} \label{genone}
w(z_j,\bar{z}_k) = \exp\left( -z_j\bar{z}_k\right) . 
\end{equation}
It is natural to expect that the statistical mean values of an infinite system 
do not depend on the gauge of the background potential, so that 
the radius-dependent potential $u(r)$ induces the uniform particle density; 
the mathematical formalism which is beyond this phenomenon was developed 
in Ref. \cite{Samaj98}.
The two-body densities are translational invariant, i.e., they depend on
the distance between the two points: 
$n^{(2)}({\bf r},{\bf r}') = n^{(2)}(\vert {\bf r}-{\bf r}'\vert)$ and
$h({\bf r},{\bf r}') = h(\vert {\bf r}-{\bf r}'\vert)$.   

The two-body density $n^{(2)}(z_1,\bar{z}_1,z_2,\bar{z}_2) 
= n^{(2)}(\vert z_1-z_2\vert)$ is at small distances
$\vert z_1-z_2\vert\to 0$ proportional to the interaction Boltzmann 
factor of the two particles $\vert z_1-z_2\vert^{2\gamma}$ and the remaining 
part of its short-distance expansion is analytic in $\vert z_1-z_2\vert^2$
\cite{Samaj95}:
\begin{equation}
n^{(2)}(\vert z_1-z_2\vert) = \left[(z_1-z_2)(\bar{z}_1-\bar{z}_2)\right]^{\gamma}
\sum_{j=0}^{\infty} a_j(\gamma) \left[(z_1-z_2)(\bar{z}_1-\bar{z}_2)\right]^j ,
\end{equation}
where the expansion coefficients $\{ a_j\}$ depend on $\gamma$.
Considering the symmetry relation (\ref{final1}) with the generalized
Boltzmann factor (\ref{genone}), one obtains the equality
\begin{eqnarray}
\sum_{j=0}^{\infty} a_j(\gamma) \left[(z_1-z_2)(\bar{z}_1-\bar{z}_2) \right]^j 
& = & {\rm e}^{-(z_1-z_2)(\bar{z}_1-\bar{z}_2)} \nonumber \\ & & \times
\sum_{j=0}^{\infty} a_j(\gamma) (-1)^j 
\left[(z_1-z_2)(\bar{z}_1-\bar{z}_2) \right]^j . \nonumber \\ & &
\end{eqnarray}
Denoting the particle distance as $r=\sqrt{(z_1-z_2)(\bar{z}_1-\bar{z}_2)}$,
the preceding two relations can be written in a simpler way as 
\begin{eqnarray}
n^{(2)}(r) & = & r^{2\gamma} \sum_{j=0}^{\infty} a_j(\gamma) r^{2j} , \label{a} \\
\sum_{j=0}^{\infty} a_j(\gamma) r^{2j} & = & {\rm e}^{-r^2} 
\sum_{j=0}^{\infty} a_j(\gamma) (-1)^j r^{2j} . \label{b}
\end{eqnarray}
Since the function $n^{(2)}({\rm i}r)$ of the purely imaginary distance
${\rm i}r$ is well defined in terms of the small-$r$ expansion as follows
\begin{equation}
n^{(2)}({\rm i}r) = ({\rm i}r)^{2\gamma} \sum_{j=0}^{\infty} a_j(\gamma) 
({\rm i}r)^{2j} ,
\end{equation}
the relation (\ref{b}) is equivalent to the one
\begin{equation}
n^{(2)}(r) = (-1)^{\gamma} {\rm e}^{-r^2} n^{(2)}({\rm i}r) . 
\end{equation}
If applied twice, it is an identity.
The analogous formula for the pair correlation function reads as
\begin{equation} \label{relationh}
h(r) = -1 + (-1)^{\gamma} {\rm e}^{-r^2} 
+ (-1)^{\gamma} {\rm e}^{-r^2} h({\rm i}r) .
\end{equation}

The relation (\ref{b}) can be reexpressed as
\begin{eqnarray}
{\rm e}^{r^2/2} \sum_{j=0}^{\infty} a_j(\gamma) r^{2j} & = & {\rm e}^{-r^2/2} 
\sum_{j=0}^{\infty} a_j(\gamma) (- r^2)^j \nonumber \\
& = & {\rm e}^{({\rm i}r)^2/2} \sum_{j=0}^{\infty} a_j(\gamma) ({\rm i}r)^{2j} .
\end{eqnarray}
Consequently,
\begin{equation}
{\rm e}^{r^2/2} \sum_{j=0}^{\infty} a_j(\gamma) r^{2j} =
\sum_{j=0}^{\infty} b_j(\gamma) r^{4j}
\end{equation}
with some new expansion coefficients $\{ b_j\}$.
We conclude that possible short-distance functional forms of $n^{(2)}(r)$ 
reduce themselves to
\begin{equation}
n^{(2)}(r) = {\rm e}^{-r^2/2} r^{2\gamma} \sum_{j=0}^{\infty} b_j(\gamma) r^{4j} . 
\end{equation}
The pair correlation function is expressible as
\begin{equation}
h(r) = - 1 + {\rm e}^{-r^2/2} r^{2\gamma} \sum_{j=0}^{\infty} c_j(\gamma) r^{4j} ,
\end{equation}
where $c_j(\gamma)=b_j(\gamma)/n^2$.
In particular, for the exactly solvable $\gamma=1$ with $h(r)=-{\rm e}^{-r^2}$
it is easy to verify that
\begin{equation} \label{c1}
c_j(1) = \frac{1}{(2j+1)!} \frac{1}{2^{2j}} .
\end{equation}

The pair correlation function $h(r)$ should vanish as $r\to\infty$, i.e.
\begin{equation}
\lim_{r\to\infty} {\rm e}^{-r^2/2} r^{2\gamma} \sum_{j=0}^{\infty} 
c_j(\gamma) r^{4j} = 1 .
\end{equation}
Since it holds that
\begin{equation}
\lim_{r\to\infty} {\rm e}^{-r^2/2} r^{2\gamma} \sum_{j=0}^{\infty} 
\frac{1}{(2j+\gamma)!} \frac{2}{2^{2j+\gamma}} r^{4j} = 1 ,
\end{equation}
the coefficients with asymptotically large indices should go to
\begin{equation}
\lim_{j\to\infty} c_j(\gamma) = \frac{1}{(2j+\gamma)! 2^{2j+\gamma-1}} .
\end{equation} 
Note that for $\gamma=1$ the $c$-coefficients (\ref{c1}) are
in fact equal to this asymptotic prediction.
We subtract from $c_j(\gamma)$ their asymptotic values in order to ensure
the series convergence of $h(r)$:
\begin{equation} \label{h}
h(r) = {\rm e}^{-r^2/2} r^{2\gamma} \sum_{j=0}^{\infty} \left[ 
c_j(\gamma) -\frac{1}{(2j+\gamma)! 2^{2j+\gamma-1}} \right] r^{4j} + f(r) , 
\end{equation}
where
\begin{equation}
f(r) = {\rm e}^{-r^2/2} r^{2\gamma} \sum_{j=0}^{\infty} 
\frac{1}{(2j+\gamma)! 2^{2j+\gamma-1}} r^{4j} - 1 .
\end{equation}
The explicit form of the function $f(r)$ depends on whether $\gamma$ is odd
or even. 
In particular, for $\gamma=2g+1$ $(g=0,1,2,\ldots)$ one has
\begin{equation} \label{f1}
f(r) = - {\rm e}^{-r^2} - {\rm e}^{-r^2/2} \sum_{j=0}^{g-1} 
\frac{1}{(2j+1)! 2^{2j}} r^{2(2j+1)}
\end{equation}
and for $\gamma=2g$ $(g=1,2,\ldots)$
\begin{equation} \label{f2}
f(r) = {\rm e}^{-r^2} - {\rm e}^{-r^2/2} \sum_{j=0}^{g-1} 
\frac{1}{(2j)! 2^{2j-1}} r^{4j} .   
\end{equation}
To simplify the notation, one defines
\begin{equation} \label{dj}
d_j(\gamma) \equiv c_j(\gamma) -\frac{1}{(2j+\gamma)! 2^{2j+\gamma-1}} .
\end{equation}

\renewcommand{\theequation}{4.\arabic{equation}}
\setcounter{equation}{0}

\section{Self-dual relation} 
\label{Sect4}
In this paper, the 2D Fourier transform $\tilde{f}({\bf q})$ 
of the function $f({\bf r})$ is defined by
\begin{equation} \label{Fourier}
f({\bf r}) =  \frac{1}{2} \int \frac{{\rm d}^2q}{2\pi} 
\tilde{f}({\bf q}) {\rm e}^{-{\rm i}{\bf q}\cdot{\bf r}} , \qquad
\tilde{f}({\bf q}) =  2 \int \frac{{\rm d}^2r}{2\pi} h({\bf r})
{\rm e}^{{\rm i}{\bf q}\cdot{\bf r}} . 
\end{equation}
The prefactors 1/2 and 2 are introduced to simplify layout of final formulae.

Our next goal is to formulate the bulk symmetry relations, derived
in the previous Sect. \ref{Sect3}, in the Fourier space.  
Using (\ref{relationh}) it holds that 
\begin{eqnarray}
{\rm e}^{q^2/4} \tilde{h}({\bf q}) & = & 2 \int \frac{{\rm d}^2r}{2\pi} 
h({\bf r}) {\rm e}^{q^2/4+{\rm i}{\bf q}\cdot{\bf r}} \nonumber \\ 
& = & 2 \int \frac{{\rm d}^2r}{2\pi} 
\left[ - {\rm e}^{r^2} + (-1)^{\gamma} + (-1)^{\gamma} h({\rm i}r) \right]
{\rm e}^{-({\bf r}-{\rm i}{\bf q}/2)^2} .
\end{eqnarray}
This expression allows us to write
\begin{equation} \label{hrepre}
{\rm e}^{q^2/4} \tilde{h}({\bf q}) = (-1)^{\gamma} + \tilde{g}({\bf q}) ,
\end{equation}
where
\begin{eqnarray} 
\tilde{g}({\bf q}) & = & 2 \int \frac{{\rm d}^2r}{2\pi}  
\left[ (-1)^{\gamma} h({\rm i}{\bf r}) - 
{\rm e}^{r^2} \right] {\rm e}^{-({\bf r}-{\rm i}{\bf q}/2)^2} \nonumber \\  
& = & 2 \int \frac{{\rm d}^2r}{2\pi}  
\left[ (-1)^{\gamma} h({\rm i}{\bf r}-{\bf q}/2) - 
{\rm e}^{({\bf r}+{\rm i}{\bf q}/2)^2} \right] {\rm e}^{-r^2} . \label{g}
\end{eqnarray}
The function $\tilde{g}(q)$ is related to the guiding-center structure 
factor $\hat{s}(q)-\hat{s}_{\infty}$ \cite{Haldane11a,Haldane11b} as follows 
$\tilde{g}(q) = \gamma^2 [\hat{s}(q)-\hat{s}_{\infty}]$. 
As $q\to\infty$, $\tilde{g}(q)$ goes evidently to 0, since 
the correlation function $h$ vanishes for large distances and also
${\rm e}^{({\bf r}+{\rm i}{\bf q}/2)^2}$ goes to 0 when $q\to\infty$.
It is evident from Eq. (\ref{hrepre}) that this fact permits one
to determine the large-$q$ asymptotic behavior of $\tilde{h}({\bf q})$
for any $\gamma=1,2,\ldots$: 
\begin{equation} \label{defasym}
\tilde{h}(q) = (-1)^{\gamma} {\rm e}^{-q^2/4} +  
o\left({\rm e}^{-q^2/4}\right) .  
\end{equation} 
For the exactly solvable $\gamma=1$ case \cite{Jancovici81}, 
which corresponds to
\begin{equation}
h(r) = - {\rm e}^{-r^2} , \qquad \tilde{h}(q) = - {\rm e}^{-q^2/4} ,
\end{equation}
we have trivially $\tilde{g}(q) = 0$.

Haldane \cite{Haldane11a,Haldane11b} has shown by using specific methods of 
quantum geometry that the guiding-center structure factor satisfies 
a self-dual relation between its Fourier component and the direct picture
in the 2D Euclidean space.
In what follows, we shall rederive this self-dual relation for 
the related $g$-function in a more direct way by using the representation 
(\ref{g}) which can be expressed as follows
\begin{equation} \label{gq}
(-1)^{\gamma} \tilde{g}({\bf q}) = I(q) -2 (-1)^{\gamma} \int 
\frac{{\rm d}^2r}{2\pi} {\rm e}^{({\bf r}+{\rm i}{\bf q}/2)^2} {\rm e}^{-r^2} , 
\end{equation}
where $I(q)$ is the integral
\begin{equation}
I(q) = 2 \int \frac{{\rm d}^2r}{2\pi} h({\rm i}{\bf r}-{\bf q}/2) 
{\rm e}^{-r^2} .
\end{equation}
This integral can be manipulated in the following way
\begin{eqnarray}
I(q) & = & \int \frac{{\rm d}^2r}{2\pi} {\rm e}^{-r^2} 
\int \frac{{\rm d}^2q'}{2\pi} \tilde{h}({\bf q}')
{\rm e}^{{\bf q'}\cdot{\bf r}+{\rm i}{\bf q}\cdot{\bf q}'/2} \nonumber \\
& = & \int \frac{{\rm d}^2q'}{2\pi} \tilde{h}({\bf q}')
{\rm e}^{{\rm i}{\bf q}\cdot{\bf q}'/2}
\int \frac{{\rm d}^2r}{2\pi} {\rm e}^{-r^2+{\bf q}'\cdot{\bf r}} \nonumber \\
& = & \int \frac{{\rm d}^2q'}{2\pi} \tilde{h}({\bf q}')
{\rm e}^{{\rm i}{\bf q}\cdot{\bf q}'/2} \frac{1}{2} {\rm e}^{q'^2/4} .
\end{eqnarray}
Inserting here the representation (\ref{hrepre}), one ends up with
\begin{equation}
I(q) = \frac{1}{2} \int \frac{{\rm d}^2q'}{2\pi} \tilde{g}({\bf q}')
{\rm e}^{{\rm i}{\bf q}\cdot{\bf q}'/2} 
+ \frac{1}{2} (-1)^{\gamma}\int \frac{{\rm d}^2q'}{2\pi} 
{\rm e}^{{\rm i}{\bf q}\cdot{\bf q}'/2} .
\end{equation} 
Considering this relation in (\ref{gq}), one gets
\begin{eqnarray} 
(-1)^{\gamma} \tilde{g}({\bf q}) & = & 
\frac{1}{2} \int \frac{{\rm d}^2q'}{2\pi} \tilde{g}({\bf q}')
{\rm e}^{{\rm i}{\bf q}\cdot{\bf q}'/2} 
+ \frac{1}{2} (-1)^{\gamma}\int \frac{{\rm d}^2q'}{2\pi} 
{\rm e}^{{\rm i}{\bf q}\cdot{\bf q}'/2} \nonumber \\
& & -2 (-1)^{\gamma} \int \frac{{\rm d}^2r}{2\pi}
{\rm e}^{{\rm i}{\bf r}\cdot{\bf q}-q^2/4} .
\end{eqnarray}
The last two terms are proportional to the Dirac $\delta(q)$ and they
cancel with one another.
Thus one arrives at the self-dual relation for the $g$-function
\begin{equation} \label{selfdual} 
(-1)^{\gamma} \tilde{g}({\bf q}) = 
\frac{1}{2} \int \frac{{\rm d}^2q'}{2\pi} \tilde{g}({\bf q}')
{\rm e}^{{\rm i}{\bf q}\cdot{\bf q}'/2} . 
\end{equation}
This self-dual formula relates the Fourier and Euclidean pictures of 
the $g$-function in the following way
\begin{equation} \label{selfdualp}
\tilde{g}(q) = (-1)^{\gamma} g(r=q/2) .
\end{equation}

Next aim is to incorporate the general symmetry formulae
(\ref{h})--(\ref{dj}) for the pair correlation function $h(r)$ into
the ones formulated in the Fourier space for the function $\tilde{g}(q)$ 
in such a way that the self-dual relation (\ref{selfdualp}) 
be automatically satisfied.
The procedure depends on whether $\gamma$ is an odd or even integer. 

\subsection{$\gamma=2g+1$} \label{Sect4.1}
Let us first treat the case $\gamma=2g+1$ $(g=1,2,\ldots)$ for which 
it holds that
\begin{eqnarray}
h(r) + {\rm e}^{-r^2} & = &
{\rm e}^{-r^2/2} \sum_{j=0}^{\infty} d_j r^{2+4(j+g)} \nonumber \\
& & - {\rm e}^{-r^2/2} \sum_{j=0}^{g-1} \frac{1}{(2j+1)! 2^{2j}} r^{2(2j+1)} .    
\end{eqnarray}
One goes to the Fourier space by writing 
${\rm d}^2r = r {\rm d}r {\rm d}\varphi$ and using the definition
of the Bessel function \cite{Gradshteyn}
\begin{equation}
\int_0^{2\pi} \frac{{\rm d}\varphi}{2\pi} {\rm e}^{{\rm i} q r \cos\varphi}
= J_0(qr) ,
\end{equation}
to obtain
\begin{eqnarray}
\tilde{h}(q) + {\rm e}^{-q^2/4} & = & \sum_{j=0}^{\infty} d_j
2^{2(1+j+g)} \int_0^{\infty} {\rm d}t\, {\rm e}^{-t} t^{1+2(j+g)} 
J_0\left( 2\sqrt{\frac{q^2}{2}t}\right) \nonumber \\ & &
- 4 \sum_{j=0}^{g-1} \frac{1}{(2j+1)!}
\int_0^{\infty} {\rm d}t\, {\rm e}^{-t} t^{1+2j} 
J_0\left( 2\sqrt{\frac{q^2}{2}t}\right) . \nonumber \\ & &
\end{eqnarray}
Applying the relation
\begin{equation}
\int_0^{\infty} {\rm d}t\, {\rm e}^{-t} t^j J_0(2\sqrt{rt}) 
= {\rm e}^{-r} n! L_j(r)
\end{equation}
with $\{ L_j(r) \}_{j=0}^{\infty}$ being the standard Laguerre polynomials
\cite{Gradshteyn}, with regard to the definition (\ref{hrepre}) one gets 
the following representation of the $g$-function:
\begin{eqnarray} \label{grepres}
\tilde{g}(q) & = & {\rm e}^{-q^2/4} \sum_{j=0}^{\infty} d_j
2^{2(1+j+g)} [1+2(j+g)]! L_{1+2(j+g)}(q^2/2) \nonumber \\ & &
- 4 {\rm e}^{-q^2/4} \sum_{j=0}^{g-1} L_{1+2j}(q^2/2) .   
\end{eqnarray}
This representation automatically fulfills the self-dual relation 
(\ref{selfdual}).
Indeed, inserting into the equality
\begin{equation} \label{self} 
\frac{1}{2} \int \frac{{\rm d}^2q'}{2\pi} \tilde{g}({\bf q}')
{\rm e}^{{\rm i}{\bf q}\cdot{\bf q}'/2} = \frac{1}{4} \int_0^{\infty} {\rm d}t\,
\tilde{g}(\sqrt{t}) J_0\left( \frac{q}{2}\sqrt{t}\right) ,
\end{equation}
the representation (\ref{grepres}) and using the formula \cite{Gradshteyn} 
\begin{equation} \label{formulaA}
\int_0^{\infty} {\rm d}t\, {\rm e}^{-t} L_{1+2n}(2t) J_0(2\sqrt{rt}) 
= - {\rm e}^{-r} L_{1+2n}(2r) 
\end{equation}
one recovers the self-dual relation (\ref{selfdual}) with $(-1)^{\gamma}=-1$.
The representation (\ref{grepres}) is in fact the most general series
representation of $\tilde{g}(q)$ which satisfies the self-dual relation.

The Laguerre polynomials satisfy the orthogonality relations \cite{Gradshteyn}
\begin{equation} \label{ortho}
\int_0^{\infty} {\rm d}r\, {\rm e}^{-r} L_n(r) L_m(r) = \delta_{nm}
= \int_0^{\infty} {\rm d}q\, q {\rm e}^{-q^2/2} L_n(q^2/2) L_m(q^2/2) . 
\end{equation}
Multiplying the representation of $\tilde{g}(q)$ (\ref{grepres})
with $q {\rm e}^{-q^2/4} L_{2j}(q^2/2)$ $(j=1,2,\ldots)$ and integrating
over $q$ from 0 to $\infty$, these orthogonality relations imply 
an infinite sequence of zero integrals
\begin{equation} \label{LagA}
\int_0^{\infty} {\rm d}q\, q {\rm e}^{-q^2/4} 
\tilde{g}(q) L_{2j}(q^2/2) = 0 \qquad \mbox{for $j=0,1,2,\ldots$.}
\end{equation}
On the other hand, the multiplication of (\ref{grepres}) with
$q {\rm e}^{-q^2/4} L_{2j+1}(q^2/2)$ $(j=1,2,\ldots)$ and the consequent
integration over $q$ implies that
\begin{eqnarray}
\int_0^{\infty} {\rm d}q\, q {\rm e}^{-q^2/4} \tilde{g}(q) L_{2j+1}(q^2/2) 
& = & -4 \qquad \mbox{for $j=0,1,\ldots,g-1$,} \nonumber \\
\int_0^{\infty} {\rm d}q\, q {\rm e}^{-q^2/4} 
\tilde{g}(q) L_{2j+1}(q^2/2) & = & d_{j-g} 2^{2(1+j)} (2j+1)! \nonumber \\
& & \phantom{-4} \qquad \mbox{for $j=g,g+1,\ldots$.}
\end{eqnarray}

\subsection{$\gamma=2g$} \label{Sect4.2}
If $\gamma=2g$ $(g=1,2,\ldots)$, one has
\begin{equation}
h(r) - {\rm e}^{-r^2} =
{\rm e}^{-r^2/2} \sum_{j=0}^{\infty} d_j r^{4(j+g)}
- {\rm e}^{-r^2/2} \sum_{j=0}^{g-1} \frac{1}{(2j)! 2^{2j-1}} r^{4j} .    
\end{equation}
The Fourier transform of this equation implies
\begin{eqnarray} \label{gprepres}
\tilde{g}(q) & = & {\rm e}^{-q^2/4} \sum_{j=0}^{\infty} d_j 2^{1+2(j+g)}
[2(j+g)]! L_{2(j+g)}(q^2/2) \nonumber \\ & &
- 4 {\rm e}^{-q^2/4} \sum_{j=0}^{g-1} L_{2j}(q^2/2) .   
\end{eqnarray}
As before, this representation automatically fulfills the self-dual relation 
(\ref{selfdual}) with $(-1)^{\gamma}=1$.
This can be shown by inserting the representation (\ref{gprepres}) into
(\ref{self}) and by using the formula \cite{Gradshteyn}
\begin{equation}
\int_0^{\infty} {\rm d}t\, {\rm e}^{-t} L_{2n}(2t) J_0(2\sqrt{rt}) 
= + {\rm e}^{-r} L_{2n}(2r) ; 
\end{equation}
note the plus sign in comparison with (\ref{formulaA}).

Multiplying the representation of $\tilde{g}(q)$ (\ref{gprepres})
with $q {\rm e}^{-q^2/4} L_{2j+1}(q^2/2)$ $(j=1,2,\ldots)$ and integrating
over $q$, the orthogonality relations (\ref{ortho}) imply 
an infinite sequence of zero integrals
\begin{equation} \label{LagB}
\int_0^{\infty} {\rm d}q\, q {\rm e}^{-q^2/4} \tilde{g}(q) L_{2j+1}(q^2/2) 
= 0 \qquad \mbox{for $j=0,1,2,\ldots$.}
\end{equation}
The multiplication of (\ref{gprepres}) with
$q {\rm e}^{-q^2/4} L_{2j}(q^2/2)$ $(j=1,2,\ldots)$ and the consequent
integration over $q$ leads to
\begin{eqnarray}
\int_0^{\infty} {\rm d}q\, q {\rm e}^{-q^2/4} \tilde{g}(q) L_{2j}(q^2/2) & = & 
-4 \qquad \mbox{for $j=0,1,\ldots,g-1$,} \nonumber \\
\int_0^{\infty} {\rm d}q\, q {\rm e}^{-q^2/4} 
\tilde{g}(q) L_{2j}(q^2/2) & = & d_{j-g} 2^{2j+1} (2j)!  \nonumber \\
& & \phantom{-4} \qquad \mbox{for $j=g,g+1,\ldots$.}
\end{eqnarray}

\renewcommand{\theequation}{5.\arabic{equation}}
\setcounter{equation}{0}

\section{Semi-infinite geometry} \label{Sect5}
Let us now consider the 2D geometry of the plain hard wall in the half-space
$x<0$ and the charged particles constrained to the complementary
half-space $x\ge 0$.
The system is infinite in the $y$ direction, $y\in (-\infty,\infty)$.
The wall surface at $x=0$ is charged by a fixed ``line'' charge density 
$-e\sigma$.  
The two-body density $n^{(2)}(z_1,\bar{z}_1,z_2,\bar{z}_2)$ is translationally
invariant along the $y$-axis, i.e. it depends on $\vert y_1-y_2\vert$.
As concerns the $x$-axis, taking into account the particle interchangeability,
in the center-of-mass basis (\ref{mass}) the two-body density depends on 
$(x_1+x_2)/2$ and $\vert x_1-x_2\vert$.
Thus
\begin{equation}
n^{(2)}(z_1,\bar{z}_1,z_2,\bar{z}_2) \equiv
n^{(2)}\left( \tfrac{x_1+x_2}{2},x_1-x_2,y_1-y_2\right) .
\end{equation}

Like in the bulk case, the two-body density is proportional to the interaction 
Boltzmann factor of the two particles $\vert z_1-z_2\vert^{2\gamma}$
when $\vert z_1-z_2\vert\to 0$. 
The remaining part of the short-distance expansion is assumed to be analytic
in small quantities $(x_1-x_2)^2$ and $(y_1-y_2)^2$:
\begin{eqnarray}
n^{(2)}\left( \tfrac{x_1+x_2}{2},x_1-x_2,y_1-y_2\right) & = &
\left[ (x_1-x_2)^2 + (y_1-y_2)^2 \right]^{\gamma} \nonumber \\
& & \times \sum_{j,k=0}^{\infty} a_{jk}\left(\tfrac{x_1+x_2}{2}\right)
(x_1-x_2)^{2j} (y_1-y_2)^{2k} . \nonumber \\ & & \label{shortexp}
\end{eqnarray}
The expansion coefficients $a_{jk}$ depend on the $x$-component of 
the center-of-mass which is not small but can be any positive number. 

Under the transformation of particle coordinates (\ref{transcoor}),
the coordinate combination
\begin{equation}
\frac{1}{2} (x_1+x_2) = \frac{1}{4} \left( z_1+\bar{z}_1+z_2+\bar{z}_2 \right)
\end{equation}
remains invariant while
\begin{eqnarray}
x_1-x_2 = \frac{1}{2} \left( z_1+\bar{z}_1-z_2-\bar{z}_2 \right)
& \to & \frac{1}{2} \left( z'_2+\bar{z}'_1-z'_1-\bar{z}'_2 \right)
= {\rm i} (y'_2-y'_1) , \nonumber \\
y_1-y_2 = \frac{1}{2{\rm i}} \left( z_1-\bar{z}_1-z_2+\bar{z}_2 \right)
& \to & \frac{1}{2{\rm i}} \left( z'_2-\bar{z}'_1-z'_1+\bar{z}'_2 \right)
= {\rm i} (x'_1-x'_2) . \nonumber \\ & &
\end{eqnarray}
The symmetry relation (\ref{final2}) then takes the form
\begin{equation} \label{final3}
\frac{n^{(2)}\left(\tfrac{x_1+x_2}{2},x_1-x_2,y_1-y_2\right)}{w(z_1,\bar{z}_1) 
w(z_2,\bar{z}_2)} = (-1)^{\gamma}
\frac{n^{(2)}\left(\tfrac{x_1+x_2}{2},{\rm i}(y_2-y_1),
{\rm i}(x_1-x_2)\right)}{w(z_2,\bar{z}_1) w(z_1,\bar{z}_2)} .
\end{equation}

Two kinds of OCP Coulomb systems will be considered. 
The first ``dense'' one is the standard jellium with the fixed bulk and
wall surface charge densities.
The second ``sparse'' model corresponds to the special case of the OCP with 
no bulk background charge density, i.e. the neutral system of counter-ions 
to the charged wall surface.
For each of these models, we start with a general theory valid for any
positive integer $\gamma$ and then verify the obtained results at the
free fermion point $\gamma=1$. 

\subsection{OCP} \label{Sect5.1}

\subsubsection{General theory}
We consider the standard 2D OCP with a fixed volume background charge 
density $-e n$ in the half-space $x>0$ and line charge density 
$-e\sigma$ at the wall surface $x=0$.
The electrostatic potential induced by the background charge density, 
given by the Poisson equation
\begin{equation}
\frac{{\rm d}^2 u(x)}{{\rm d}x^2} = 2\pi n e ,
\end{equation}
reads as $u_1(x)=\pi n e x^2$.
The potential induced by the line charge density is $u_2(x)=\pi \sigma e x$.
The corresponding one-body Boltzmann factor is
\begin{equation}
w(z,\bar{z}) = \exp\left[ -\Gamma \pi n \left( \tfrac{z+\bar{z}}{2} \right)^2 
- \Gamma\pi\sigma \left( \tfrac{z+\bar{z}}{2} \right) \right] . 
\end{equation}
Expressing all distances in units of length $a$ (\ref{lengtha}), 
the generalized Boltzmann factor reads as
\begin{equation}
w(z_j,\bar{z}_k) = \exp\left[ - \frac{1}{2} \left( z_j+\bar{z}_k \right)^2 
- \gamma \pi \sigma \left( z_j+\bar{z}_k \right) \right] , 
\end{equation}
where $\sigma$ is the dimensionless line charge density
$\sigma/\sqrt{\gamma\pi n}$, in units of $\gamma\pi n=1$. 

The symmetry formula (\ref{final2}) leads to the relation
\begin{equation} \label{finalfinal}
n^{(2)}(z_1,\bar{z}_1,z_2,\bar{z}_2) = (-1)^{\gamma} {\rm e}^{-\vert z_1-z_2\vert^2}
n^{(2)}(z_2,\bar{z}_1,z_1,\bar{z}_2) .
\end{equation}
The equivalent formula (\ref{final3}) implies that 
\begin{eqnarray} 
n^{(2)}\left(\tfrac{x_1+x_2}{2},x_1-x_2,y_1-y_2\right)
& = & (-1)^{\gamma} {\rm e}^{-\left[ (x_1-x_2)^2+(y_1-y_2)^2 \right]} \nonumber \\
& & \times
n^{(2)}\left(\tfrac{x_1+x_2}{2},{\rm i}(y_2-y_1),{\rm i}(x_1-x_2)\right) .
\phantom{aaaa} \label{final4}
\end{eqnarray}
Note that the last two equations do not depend explicitly on $\sigma$.
Introducing the auxiliary function
\begin{eqnarray}
f\left(\tfrac{x_1+x_2}{2},x_1-x_2,y_1-y_2\right)
& = & {\rm e}^{\frac{1}{2}\left[ (x_1-x_2)^2+(y_1-y_2)^2 \right]} \nonumber \\
& & \times
n^{(2)}\left(\tfrac{x_1+x_2}{2},{\rm i}(y_2-y_1),{\rm i}(x_1-x_2)\right) ,
\phantom{aaa} \label{ff}
\end{eqnarray}
the symmetry relation (\ref{final4}) can be rewritten as
\begin{equation} \label{final5}
f\left(\tfrac{x_1+x_2}{2},x_1-x_2,y_1-y_2\right) = (-1)^{\gamma} 
f\left(\tfrac{x_1+x_2}{2},{\rm i}(y_2-y_1),{\rm i}(x_1-x_2)\right) .
\end{equation}

The short-distance expansion of the two-body density (\ref{shortexp}) 
leads to a similar expansion for the $f$-function:
\begin{eqnarray}
f\left( \tfrac{x_1+x_2}{2},x_1-x_2,y_1-y_2\right) & = &
\left[ (x_1-x_2)^2 + (y_1-y_2)^2 \right]^{\gamma} \nonumber \\
& & \times \sum_{j,k=0}^{\infty} b_{jk}\left(\tfrac{x_1+x_2}{2}\right)
(x_1-x_2)^{2j} (y_1-y_2)^{2k}  \nonumber \\ & & \label{fexp} 
\end{eqnarray}
with some other expansion coefficients $b_{jk}$ which depend on 
the $x$-component of the center-of-mass of the two particles. 
Eq. (\ref{final5}) then implies the following relation
between the coefficients $b_{jk}$: 
\begin{equation} \label{finsym}
b_{jk}\left(\tfrac{x_1+x_2}{2}\right) = (-1)^{j+k}
b_{kj}\left(\tfrac{x_1+x_2}{2}\right) .
\end{equation}  
This symmetry has no effect on diagonal coefficients $b_{jj}$
and reduces the number of independent off-diagonal coefficients
$b_{jk}$ $(j\ne k)$ by two.
Thus the most general form of the two-body density which accounts
for the present symmetry reads as
\begin{eqnarray}
n^{(2)}\left( \tfrac{x_1+x_2}{2},x_1-x_2,y_1-y_2\right) & = &
{\rm e}^{-\frac{1}{2}\left[ (x_1-x_2)^2+(y_1-y_2)^2 \right]} \nonumber \\
& & \times
\left[ (x_1-x_2)^2 + (y_1-y_2)^2 \right]^{\gamma} \nonumber \\
& & \times \bigg\{ \sum_j b_{jj}\left(\tfrac{x_1+x_2}{2}\right)
\left[(x_1-x_2)^2 (y_1-y_2)^2\right]^j  \nonumber \\ & & 
+ \sum_{j<k} b_{jk}\left(\tfrac{x_1+x_2}{2}\right) \left[
(x_1-x_2)^{2j} (y_1-y_2)^{2k} \right. \nonumber \\ & & \left.
+ (-1)^{j+k} (x_1-x_2)^{2k} (y_1-y_2)^{2j} \right] \bigg\} ,
\end{eqnarray}
where the summation indices run over integers from 0 to $\infty$. 

\subsubsection{Free-fermion point}
The semi-infinite 2D OCP at the free-fermion coupling $\Gamma=2$
was solved by Jancovici \cite{Jancovici82}.
In units of $\pi n=1$, the particle density profile was obtained in the form
\begin{equation} \label{dens}
\frac{n(x)}{n} = \frac{2}{\sqrt{\pi}} \int_{-\pi\sigma\sqrt{2}}^{\infty} 
\frac{{\rm d}t}{1+\phi(t)} {\rm e}^{-(t-x\sqrt{2})^2} ,
\end{equation} 
where
\begin{equation}
\phi(t) = \frac{2}{\sqrt{\pi}} \int_0^t {\rm d}u {\rm e}^{-u^2} 
\end{equation}
is the error function \cite{Gradshteyn}.
The function
\begin{equation} 
\frac{n(z_j,\bar{z}_k)}{n} = \frac{2}{\sqrt{\pi}} 
\int_{-\pi\sigma\sqrt{2}}^{\infty} \frac{{\rm d}t}{1+\phi(t)} 
{\rm e}^{-\left(t-\frac{z_j+\bar{z}_k}{\sqrt{2}}\right)^2} 
\end{equation} 
which is a two-point generalization of the density function (\ref{dens}),
$n(x) = n(z,\bar{z})$.
The two-body density is then expressible as \cite{Jancovici82}
\begin{equation} \label{twobody2}
n^{(2)}(z_1,\bar{z}_1,z_2,\bar{z}_2) = n(z_1,\bar{z}_1) n(z_2,\bar{z}_2)
- {\rm e}^{-\vert z_1-z_2\vert^2} n(z_2,\bar{z}_1) n(z_1,\bar{z}_2) .
\end{equation}

The explicit result (\ref{twobody2}) implies that
\begin{equation} 
n^{(2)}(z_2,\bar{z}_1,z_1,\bar{z}_2) = n(z_2,\bar{z}_1) n(z_1,\bar{z}_2)
- {\rm e}^{\vert z_1-z_2\vert^2} n(z_1,\bar{z}_1) n(z_2,\bar{z}_2) .
\end{equation}
The symmetry relation (\ref{finalfinal}) then evidently holds.
The auxiliary function (\ref{ff}) is expressible as 
\begin{eqnarray}
f\left(\tfrac{x_1+x_2}{2},x_1-x_2,y_1-y_2\right)
& = & \frac{(2n)^2}{\pi} {\rm e}^{\frac{1}{2}\left[-(x_1-x_2)^2+(y_1-y_2)^2\right]} 
\nonumber \\ & & \times 
\int_{-\pi\sigma\sqrt{2}}^{\infty} \frac{{\rm d}t}{1+\phi(t)} 
{\rm e}^{-\left(t-\frac{x_1+x_2}{\sqrt{2}}\right)^2} \nonumber \\ 
& & \times \int_{-\pi\sigma\sqrt{2}}^{\infty} \frac{{\rm d}s}{1+\phi(s)} 
{\rm e}^{-\left(s-\frac{x_1+x_2}{\sqrt{2}}\right)^2} \nonumber \\ 
& & \times \sum_{j=0}^{\infty} \frac{2^j(t-s)^{2j}}{(2j)!} 
\left[ (x_1-x_2)^{2j} \right. \nonumber \\ & & \left.
\qquad - (-1)^j (y_1-y_2)^{2j} \right] ,
\end{eqnarray}
where the summands with odd powers of $(t-s)$ disappear as a result of
the $t\leftrightarrow s$ symmetry of the kernel. 
Using that
\begin{eqnarray}
(x_1-x_2)^{2j} - (-1)^j (y_1-y_2)^{2j} & = &
\left[ (x_1-x_2)^2 + (y_1-y_2)^2 \right] \nonumber \\ & &
\times \sum_{k=0}^{j-1} (-1)^k (y_1-y_2)^{2k} (x_1-x_2)^{2(j-1-k)} , 
\label{formulaa} \nonumber \\ & &
\end{eqnarray}
the coefficients $b_{jk}$ of the expansion (\ref{fexp}) are 
expressible as 
\begin{eqnarray}
b_{jk}\left(\tfrac{x_1+x_2}{2}\right) & =  & 
\frac{(2n)^2}{\pi} \int_{-\pi\sigma\sqrt{2}}^{\infty} 
\frac{{\rm d}t}{1+\phi(t)} {\rm e}^{-\left(t-\frac{x_1+x_2}{\sqrt{2}}\right)^2} 
\nonumber \\ & & \times
\int_{-\pi\sigma\sqrt{2}}^{\infty} \frac{{\rm d}s}{1+\phi(s)} 
{\rm e}^{-\left(s-\frac{x_1+x_2}{\sqrt{2}}\right)^2} c_{jk}(t-s) ,
\end{eqnarray}
where
\begin{equation}
c_{jk}(t-s) = \frac{(-1)^j}{2^{j+k}} \sum_{l=0}^j \sum_{m=0}^k
\frac{(-1)^{l+m} 2^{2(l+m)+1}(t-s)^{2(l+m+1)}}{[2(l+m+1)]! (j-l)! (k-m)!} .
\end{equation}
The consequent symmetry
\begin{equation}
(-1)^k c_{jk}(t-s) = (-1)^j c_{kj}(t-s) 
\end{equation}
implies the same symmetry relation for the coefficients 
$b_{jk}\left(\tfrac{x_1+x_2}{2}\right)$ which is in agreement
with the general result (\ref{finsym}).

\subsection{Counter-ions only} \label{Sect5.2}

\subsubsection{General theory}
Let us now consider a version of the 2D OCP with zero volume background 
charge density $-e n = 0$.
As before, the particles possess the charge $e$ and therefore
they are ``counter-ions'' to the opposite charge line charge density 
$-e\sigma$ at the wall surface $x=0$.
The one-body Boltzmann factor of mobile particles is
$w(x)=\exp(-\Gamma\pi\sigma x)$.
Introducing the generalized Boltzmann factor
\begin{equation}
w(z_j,\bar{z}_k) = \exp\left[ - \gamma \pi \sigma 
\left( z_j+\bar{z}_k \right) \right] , 
\end{equation}
the symmetry formula (\ref{final2}) implies that
\begin{equation} \label{fifi}
n^{(2)}(z_1,\bar{z}_1,z_2,\bar{z}_2) = (-1)^{\gamma} 
n^{(2)}(z_2,\bar{z}_1,z_1,\bar{z}_2) 
\end{equation}
or, equivalently,
\begin{equation} \label{rrl}
n^{(2)}\left(\tfrac{x_1+x_2}{2},x_1-x_2,y_1-y_2\right) =  (-1)^{\gamma} 
n^{(2)}\left(\tfrac{x_1+x_2}{2},{\rm i}(y_2-y_1),{\rm i}(x_1-x_2)\right) .
\end{equation}

The short-distance expansion of the two-body density 
is still of type (\ref{shortexp}):
\begin{eqnarray}
n^{(2)}\left( \tfrac{x_1+x_2}{2},x_1-x_2,y_1-y_2\right) & = &
\left[ (x_1-x_2)^2 + (y_1-y_2)^2 \right]^{\gamma} \nonumber \\
& & \times \sum_{j,k=0}^{\infty} a_{jk}\left(\tfrac{x_1+x_2}{2}\right)
(x_1-x_2)^{2j} (y_1-y_2)^{2k} . \nonumber \\ & & \label{shortexpp} 
\end{eqnarray}
Inserting this expansion into Eq. (\ref{rrl}) leads to the following
symmetry relation between the expansion coefficients
\begin{equation} \label{finsymaa}
a_{jk}\left(\tfrac{x_1+x_2}{2}\right) = (-1)^{j+k}
a_{kj}\left(\tfrac{x_1+x_2}{2}\right) .
\end{equation}  
Consequently, the most general form of the two-body density reads as
\begin{eqnarray}
n^{(2)}\left( \tfrac{x_1+x_2}{2},x_1-x_2,y_1-y_2\right) & = &
\left[ (x_1-x_2)^2 + (y_1-y_2)^2 \right]^{\gamma} \nonumber \\
& & \times \bigg\{ \sum_j a_{jj}\left(\tfrac{x_1+x_2}{2}\right)
\left[(x_1-x_2)^2 (y_1-y_2)^2\right]^j  \nonumber \\ & & 
+ \sum_{j<k} a_{jk}\left(\tfrac{x_1+x_2}{2}\right) \left[
(x_1-x_2)^{2j} (y_1-y_2)^{2k} \right. \nonumber \\ & & \left.
+ (-1)^{j+k} (x_1-x_2)^{2k} (y_1-y_2)^{2j} \right] \bigg\} .
\end{eqnarray}

\subsubsection{Free-fermion point}
The 2D model of the charged wall with counter-ions only was solved at 
the free-fermion coupling $\Gamma=2$ in Ref. \cite{Jancovici84}.
The particle density profile, obtained in the form
\begin{equation} \label{denss}
n(x) = \frac{1}{4\pi} \int_0^{4\pi\sigma} {\rm d}s\, s {\rm e}^{-sx} ,
\end{equation}
evidently fulfills the electroneutrality condition
\begin{equation}
\int_0^{\infty} {\rm d}x n(x) = \sigma .
\end{equation}
Introducing a generalization of the density function (\ref{denss})
\begin{equation} 
n(z_j,\bar{z}_k) = \frac{1}{4\pi} \int_0^{4\pi\sigma} {\rm d}s\, s 
{\rm e}^{-s(z_j+\bar{z}_k)/2} ,
\end{equation} 
the two-body density is expressible as \cite{Jancovici84}
\begin{equation} \label{twobody22}
n^{(2)}(z_1,\bar{z}_1,z_2,\bar{z}_2) = n(z_1,\bar{z}_1) n(z_2,\bar{z}_2)
- n(z_2,\bar{z}_1) n(z_1,\bar{z}_2) .
\end{equation}

Since according to (\ref{twobody22}) one has
\begin{equation} 
n^{(2)}(z_2,\bar{z}_1,z_1,\bar{z}_2) = n(z_2,\bar{z}_1) n(z_1,\bar{z}_2)
- n(z_1,\bar{z}_1) n(z_2,\bar{z}_2) ,
\end{equation}
the symmetry relation (\ref{rrl}) holds.
The two-body density is expressible as
\begin{eqnarray} 
n^{(2)}(z_1,\bar{z}_1,z_2,\bar{z}_2) & = & \frac{1}{(4\pi)^2}
\int_0^{4\pi\sigma} {\rm d}s\, s {\rm e}^{-s\left(\frac{x_1+x_2}{2}\right)} 
\int_0^{4\pi\sigma} {\rm d}t\, t {\rm e}^{-t\left(\frac{x_1+x_2}{2}\right)} \nonumber \\
& & \times \sum_{j=0}^{\infty} \frac{(t-s)^{2j}}{2^{2j}(2j)!}
\left[ (x_1-x_2)^{2j} - (-1)^j (y_1-y_2)^{2j} \right] . \phantom{aaaa}
\end{eqnarray}
Using the relation (\ref{formulaa}), the coefficients $a_{jk}$ of 
the short-distance expansion (\ref{shortexpp}) are found to be
\begin{equation}
a_{jk}\left(\tfrac{x_1+x_2}{2}\right) = \frac{1}{(4\pi)^2}
\int_0^{4\pi\sigma} {\rm d}s\, s {\rm e}^{-s\left(\frac{x_1+x_2}{2}\right)} 
\int_0^{4\pi\sigma} {\rm d}t\, t {\rm e}^{-t\left(\frac{x_1+x_2}{2}\right)} 
b_{jk}(t-s) ,
\end{equation}
where
\begin{equation}
b_{jk}(t-s) = (-1)^k \frac{(t-s)^{2(j+k+1)}}{2^{2(j+k+1)}[2(j+k+1)]!} .
\end{equation}
Due to the equality
\begin{equation}
(-1)^j b_{jk}(t-s) = (-1)^k b_{kj}(t-s) , 
\end{equation}
the symmetry formula (\ref{finsymaa}) automatically takes place.

\renewcommand{\theequation}{6.\arabic{equation}}
\setcounter{equation}{0}

\section{Conclusion} \label{Sect6}
The studied short-distance symmetries of two-body densities for infinite 
and semi-infinite 2D OCP represent a rare occasion to get exact results 
not only at the free-fermion coupling constant $\Gamma=2$, but also at 
a sequence of couplings $\Gamma=2\times{\rm integer}$, up to 
the fluid-crystal phase transition. 

The guiding-center structure factor in the quantum Hall effect is proportional
to a specific part $\tilde{g}({\bf q})$ of the Fourier transform of the pair
correlation function of the bulk plasma $\tilde{h}({\bf q})$, given by
the relation (\ref{hrepre}).
The guiding-center factors satisfies a self-dual formula between its 
real space (Euclidean) and Fourier components \cite{Haldane11a,Haldane11b}.
The first aim of this paper was to derive this self-dual formula directly
in the format of the 2D OCP by using the short-distance symmetry of 
the pair correlation, see Eqs. (\ref{selfdual}) and (\ref{selfdualp}). 
As a by-product of the formalism, the large-$q$ asymptotic behavior of 
$\tilde{h}(q)$ (\ref{defasym}) was obtained. 
An infinite sequence of zero integrals over the Fourier component
$\tilde{g}(q)$ multiplied by Laguerre polynomials of argument $q^2/2$
was found: see Eq. (\ref{LagA}) for $\Gamma=2\times{\rm odd\ integer}$ 
and Eq. (\ref{LagB}) for $\Gamma=2\times{\rm even\ integer}$.

The second aim was to extend the short-distance symmetry of the
pair correlation function to the semi-infinite 2D OCP.
This was done for the jellium model in Sect. \ref{Sect5.1} and
for its simplified version with zero background charge density 
in Sect. \ref{Sect5.2}.
In both cases, the coefficients of the expansion in variables $(x_1-x_2)^2$ 
and $(y_1-y_2)^2$ exhibit a symmetry with respect to the permutation of 
summation indices of type (\ref{finsym}) and (\ref{finsymaa}).

The short-distance symmetry is not sufficient for determining explicitly 
the pair correlation function, however, it restricts substantially 
its possible forms. 
This might be useful in searching for the exact solution of the
2D OCP at the coupling constants $\Gamma=2\times{\rm integer}$, e.g. 
in the spirit of the work \cite{Samaj04}.

\begin{acknowledgements}
I am grateful to Prof. Duncan Haldane for pointing out my attention to the
guiding-center structure function within the fractional quantum Hall fluids.
The support received from the project EXSES APVV-16-0186 and VEGA Grant
No. 2/0003/18 is acknowledged.
\end{acknowledgements}


\begin{thebibliography}{10}

\bibitem{Alastuey81} Alastuey, A., Jancovici, B.:
On the classical two-dimensional one-component Coulomb plasma.
J. Physique {\bf 42}, 1--12 (1981)

\bibitem{Baus78} Baus, M.:
On the compressibility of a one-component plasma.
J. Phys. A: Math. Gen. {\bf 11}, 2451--2462 (1978)

\bibitem{Baus80} Baus, M., Hansen J.P.:
Statistical mechanics of simple Coulomb systems.
Phys. Rep. {\bf 59}, 1--94 (1980)

\bibitem{DiFrancesco94} Di Francesco, P., Gaudin, M., Itzykson, C., Lesage, F.:
Laughlin's wave functions, Coulomb gases and expansions of the discriminant.
Int. J. Mod. Phys. A {\bf 9}, 4257--4351 (1994)

\bibitem{Forrester98} Forrester, P.J.:
Exact results for two-dimensional Coulomb systems.
Phys. Rep. {\bf 301}, 235--270 (1998)

\bibitem{Gradshteyn} Gradshteyn, I.S., Ryzhik, I.M.:
Table of Integrals, Series and Products, 5th. edn.
Academic Press, London (1994)

\bibitem{Haldane11a} Haldane, F.D.M.:
Geometrical description of the fractional quantum Hall effect.
Phys. Rev. Lett. {\bf 107}, 116801 (2011)

\bibitem{Haldane11b} Haldane, F.D.M.:
Self-duality and long-wavelength behavior of the Landau-level guiding-center
structure function, and the shear modulus of fractional quantum Hall fluids.
arXiv:1112.0990v2 (2011)

\bibitem{Jancovici77} Jancovici, B.:
Pair correlation function in a dense plasma and pycnonuclear reactions in stars.
J. Stat. Phys. {\bf 17}, 357--370 (1977) 

\bibitem{Jancovici81} Jancovici, B.:
Exact results for the two-dimensional one-component plasma.
Phys. Rev. Lett. {\bf 46}, 386--388 (1981)

\bibitem{Jancovici82} Jancovici, B.:
Classical Coulomb systems near a plane wall. I.
J. Stat. Phys. {\bf 28}, 43--65 (1982)

\bibitem{Jancovici84} Jancovici, B.:
Surface properties of a classical two-dimensional one-component plasma:
exact results.
J. Stat. Phys. {\bf 34}, 803--815 (1984)

\bibitem{Jancovici92} Jancovici, B.:
Inhomogeneous two-dimensional plasmas.
In: Henderson. D. (ed.) Inhomogeneous Fluids, pp. 201--237, Dekker, 
New York (1992)

\bibitem{Kalinay00} Kalinay, P., Marko\v{s}, P., \v{S}amaj, L., 
Trav\v{e}nec, I.:
The sixth-moment sum rule for the pair correlations of the two-dimensional
one-component plasma: Exact result.
J. Stat. Phys. {\bf 98}, 639--666 (2000)

\bibitem{Martin88} Martin, Ph.A.:
Sum rules in charged fluids.
Rev. Mod. Phys. {\bf 60}, 1075--1127 (1988)

\bibitem{Prange87} Prange, R.E., Girvin, S.M.:
The Quantum Hall Effect, Springer, New York, (1987)

\bibitem{Samaj04} \v{S}amaj, L.:
Is the two-dimensional one-component plasma exactly solvable?
J. Stat. Phys. {\bf 117}, 131--158 (2004)

\bibitem{Samaj07} \v{S}amaj, L.:
A generalization of the Stillinger-Lovett sum rules for the two-dimensional
jellium.
J. Stat. Phys. {\bf 128}, 1415--1428 (2007)

\bibitem{Samaj95} \v{S}amaj, L., Percus, J.K.:
A functional relation among the pair correlations of the two-dimensional
one-component plasma.
J. Stat. Phys. {\bf 80}, 811--824 (1995)

\bibitem{Samaj98} \v{S}amaj, L., Kalinay, P., Trav\v{e}nec, I. :
An invariant structure of the multi-particle correlations of the
two-dimensional one-component plasma.
J. Phys. A: Math. Gen. {\bf 31}, 4149--4166 (1998)

\bibitem{Stillinger68a} Stillinger, F.H., Lovett, R.:
Ion-pair theory of concentrated electrolytes. I. Basic Concepts.
J. Chem. Phys. {\bf 48}, 3858 (1968)

\bibitem{Stillinger68b} Stillinger, F.H., Lovett, R.:
General restriction on the distribution of ions in electrolytes.
J. Chem. Phys. {\bf 49}, 1991 (1968)

\bibitem{Vieillefosse75} Vieillefosse, P., Hansen, J.P.:
Statistical mechanics of dense ionized matter. V. Hydrodynamic limit and 
transport coefficients of the classical one-component plasma.
Phys. Rev. A {\bf 12}, 1106--1116 (1975)

\bibitem{Vieillefosse85} Vieillefosse, P.:     
Sum rules and perfect screening conditions for the one-component plasma
J. Stat. Phys. {\bf 41}, 1015--1035 (1985)

\end{thebibliography}
\end{document}